\documentclass{article}

\usepackage{arxiv}

\usepackage[utf8]{inputenc} 
\usepackage[T1]{fontenc}    
\usepackage{hyperref}       
\usepackage{url}            
\usepackage{booktabs}       
\usepackage{amsfonts}       
\usepackage{nicefrac}       
\usepackage{microtype}      
\usepackage{lipsum}
\usepackage{graphicx}
\usepackage{subfigure}
\graphicspath{ {./images/} }
\usepackage{amsmath}
\usepackage{multicol}
\usepackage{multirow}

\title{Block-wise quantum grayscale image representation and compression scheme using state connection}

\author{
Md Ershadul Haque \\
  School of Computing, Mathematics and Engineering\\
  Charles Sturt University\\
  Bathurst, NSW 2795 \\
  \texttt{mhaque@csu.edu.au} \\
   \And
Manoranjan Paul \\
  School of Computing, Mathematics and Engineering\\
  Charles Sturt University\\
  Bathurst, NSW 2795 \\
  \texttt{mpaul@csu.edu.au} \\
  \And
 Anwaar Ulhaq \\
  School of Computing, Mathematics and Engineering\\
  Charles Sturt University\\
  Bathurst, NSW 2795 \\
  \texttt{aulhaq@csu.edu.au} \\
    \And
Tanmoy Debnath \\
  School of Computing, Mathematics and Engineering\\
  Charles Sturt University\\
  Bathurst, NSW 2795 \\
  \texttt{tdebnath@csu.edu.au} \\
}
\begin{document}
\maketitle
\begin{abstract}
Quantum computing draws huge attention due to its faster computational capability compared to classical computing to represent and compress the classical image data into the quantum domain. The main idea of quantum domain representation is to convert pixel intensities and their coordinates i.e. state label preparation using quantum bits i.e. Qubits. For a bigger size image, the state label preparation takes more Qubits. To address more Qubits issues, a novel SCMNEQR (State Connection Modification Novel Enhanced Quantum Representation) approach has been proposed that uses fewer qubits to map the arbitrary size of the grayscale image using block-wise state label preparation. The proposed SCMNEQR approach introduces the state connection using a reset gate rather than repeating the use of the Toffoli gate used in the existing approach. The experimental results show that the proposed approach outperforms the existing methods in terms of compression.  
\end{abstract}
\keywords{Quantum image representation and compression, image preparation, SCMNEQR, state connection}

\section{Introduction}\label{sec:intro}
In the quantum computer discipline, physics, and computer science plays an important role in concrete the quantum information processing (QIP) field \cite{khan2019}. Superposition, Entanglement, and Parallelism are the main properties in quantum computing that make it unique \cite{jacobs1963fine, ladd2010quantum,mandra2016faster, itrelease, aaronson2008limits}. On the other hand, the classical computer is unable to solve the NP (non-deterministic polynomial) hard problems rapidly. Meanwhile, its computing power has not increased significantly in the past decade \cite{wang2021review}. Therefore, Feynman et al. demonstrated another way to increase computing power by introducing the quantum computer \cite{feynman2018simulating}. An algorithm for factorial calculation of integer was proposed in \cite{shor1994algorithms}. After that, following factorial calculation, a database search algorithm was found in \cite{grover1996fast}. Many other applications, such as flight schedules, and chess playing, it’s still suffering from limited algorithmic support like today's classical computer. The ideas from the quantum computer should be on an atomic scale like classical computers \cite{itrelease,  aaronson2008limits, haque2022novel,forbes2020}. 
Applications of a quantum computer in terms of image computing are compression, segmentation, security, watermarking, quantum machine learning, and remote sensing \cite{itrelease, kalonia2021review}. The complexity of the classical computer requires $O(n*2^n)$ whereas the quantum computer is $O(n)$. In a quantum computer, the number of required operational gates that determine the complexity of the system \cite{yan2016survey}. The preparation of the grayscale image requires different strategies compared to the color counterpart.  
In this work, a novel SCMNEQR approach has been proposed for JPEG grayscale image representation and compression to make the system more practical to use. The DCT is used as a preparation step. The main contributions of this proposed work are outlined as follows: 
\begin{itemize}
    \item Representing and compressing the grayscale image in a quantum block-wise system
    \item Minimizing the bit rate tremendously compared to the previous approach 
    \item Measuring the corresponding rate-distortion curve in terms of the quantum domain 
    \item Introducing a novel reset gate in replace of the Toffoli gate 
\end{itemize}

The remaining paper is organized as follows. Section \ref{L_R} outlines the related works. Section \ref{P_M} discusses the proposed approach. Section \ref{R_D} represents the computational result and its summary. Section \ref{CC} concludes the proposed work.
\section{Related Works}\label{L_R}
\begin{figure}[htbp]
\centerline{\includegraphics[width=\linewidth]{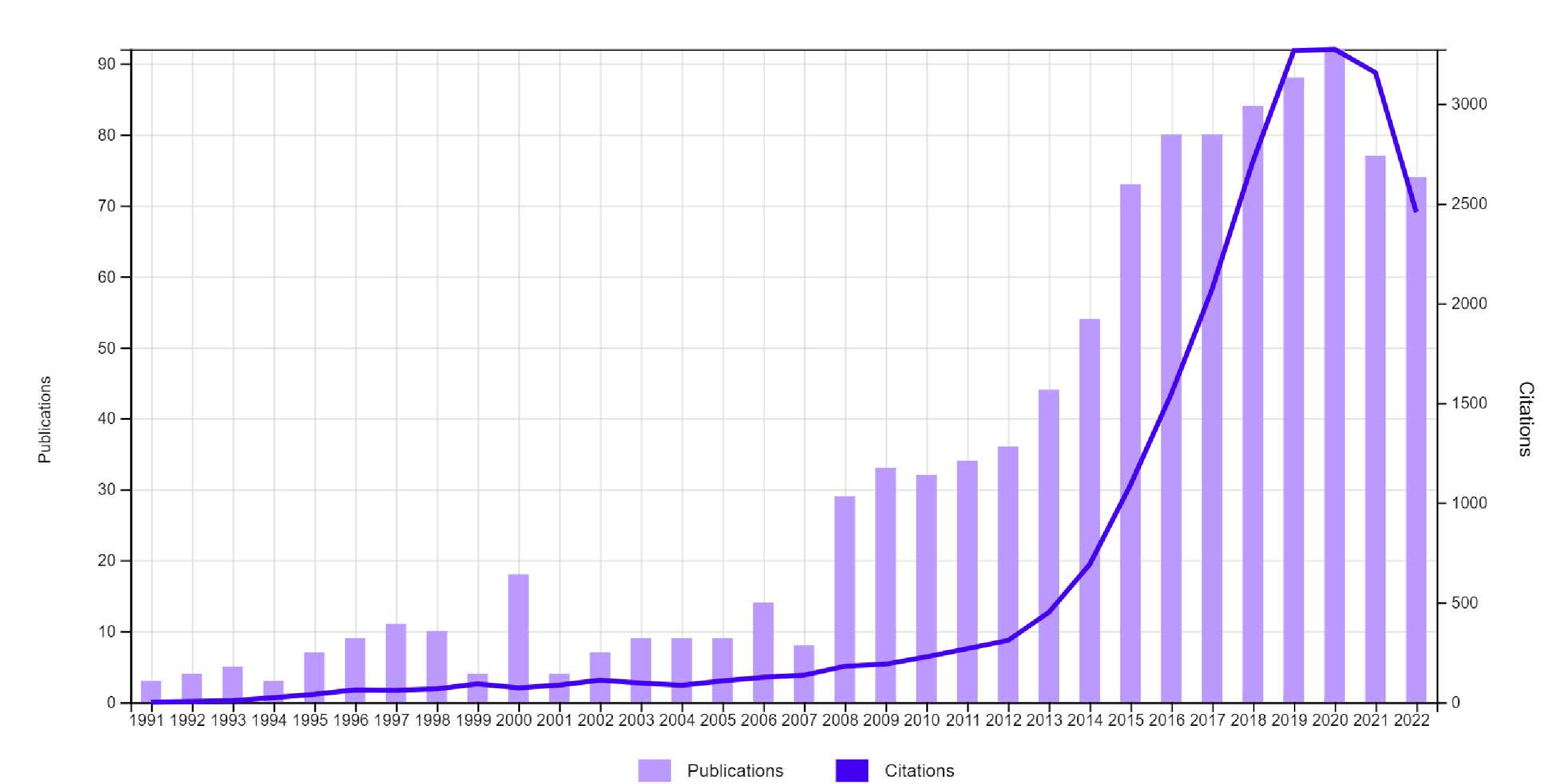}}
\caption{Number of publications and citations for quantum image representation.}
\label{fig_ls}
\end{figure}

\begin{figure}[htbp]
\centerline{\includegraphics[width=\linewidth]{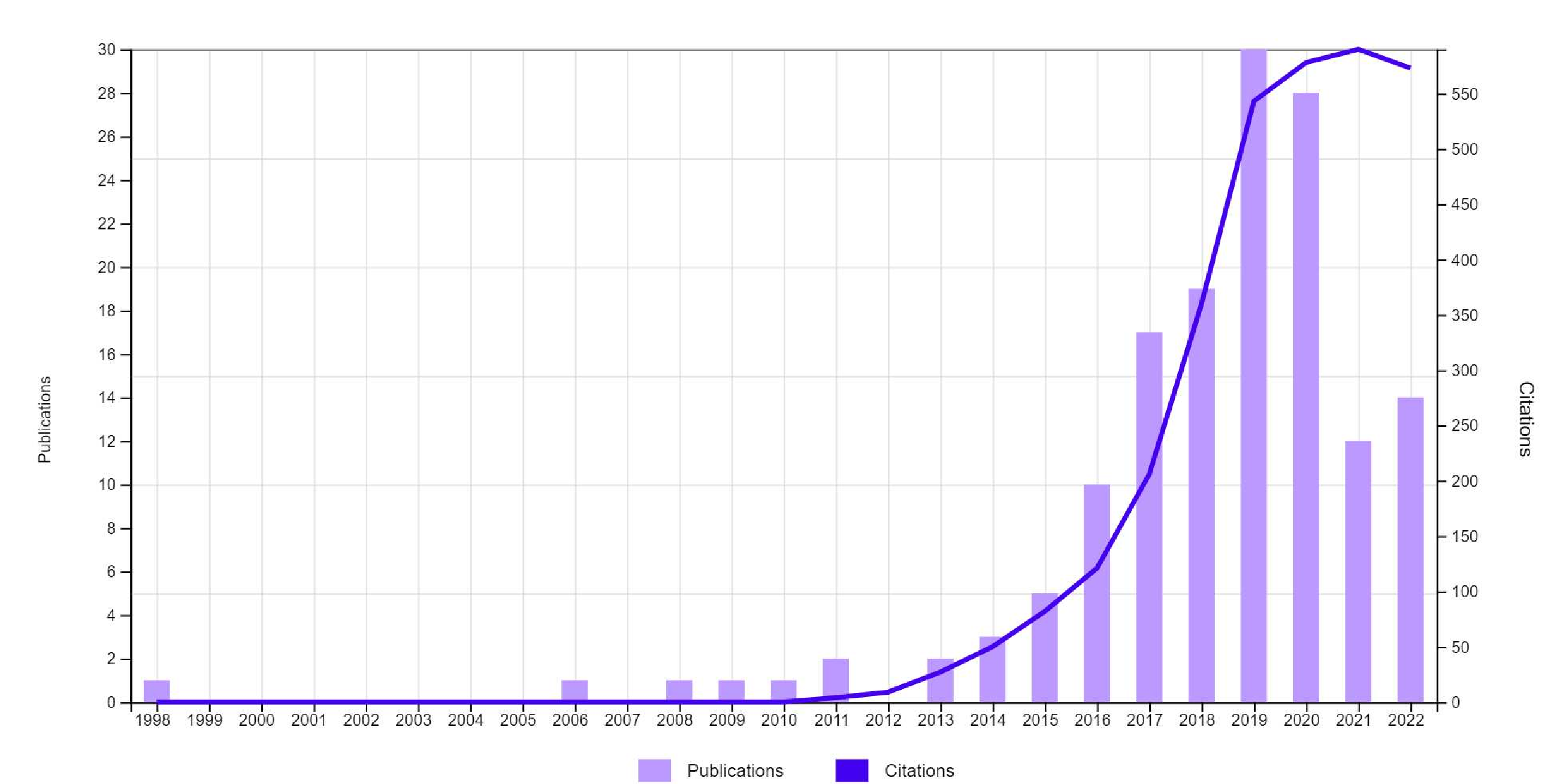}}
\caption{Number of publications and citations for quantum image representation and compression.}
\label{fig_lsc}
\end{figure}

\begin{figure}[htbp]
\centerline{\includegraphics[width=\linewidth]{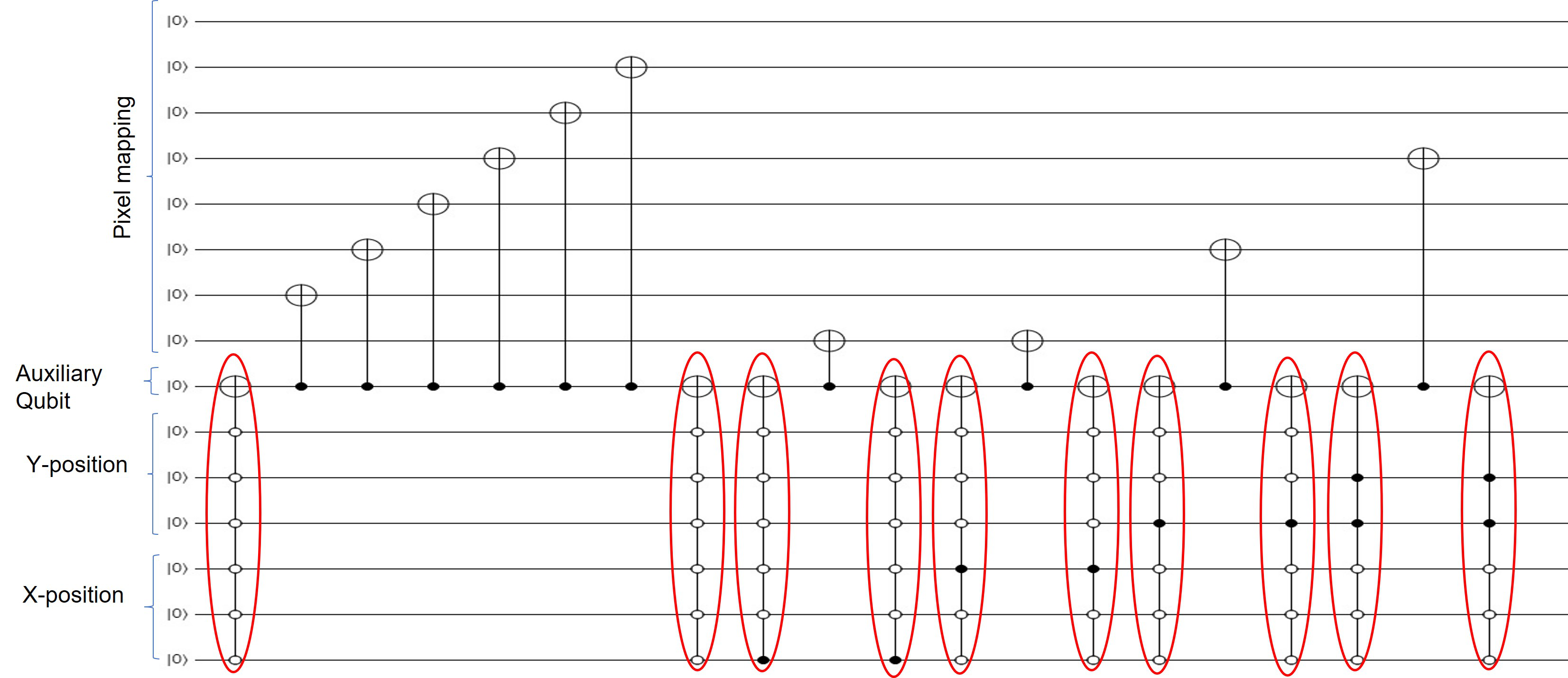}}
\caption{An EFRQI circuit diagram for pixel values representation}
\label{fig_efrqi}
\end{figure}

Figure \ref{fig_ls} shows the frequency of publications and citations on quantum image representation published between 1991 and 2022. Interest in this field is gradually increasing and the publication records were found higher in 2022 compared to other years.
Figure \ref{fig_lsc} shows the publications and citations trend for more than three decades for image representation and compression. The graph shows that the initial work of quantum representation and its compression was seen in 1998. After that, there no significant work was done until 2011. The publication report increased exponentially till 2019. Between 2019 and 2021, the number of publications changes randomly. The solid line in Figure \ref{fig_lsc} shows the citation report for image representation and compression.  The citation report started after 2011, and thereafter it grew tremendously. The citation report indicates that quantum computing is a growing field in the research community. However, still there have a lot of areas of development issues that are the main motivation of this research. 
\\
Qubit lattice is the first quantum approach that demonstrates how to represent and restore a classical image into a quantum system \cite{zhang2013neqr}. A Real-ket-based image representation algorithm was proposed by Latorre et al. in \cite{latorre2005image}. An FRQI (Flexible Representation of Quantum Image) algorithm was developed in \cite{le2011flexible} that converts four random pixels into its equivalent quantum system using a single ancillary qubit as an angle. The Entanglement image representation was proposed in 2010 \cite{venegas2010processing}. JPEG-based image compression was proposed by Jiang et al. in 2017 \cite{jiang2018novel} that uses GQIR and DCT approaches together. An equivalent bit pixel image is investigated by Laurel et al. in \cite{laurel2015equivalence}. A NEQR (Novel Enhanced Quantum Representation), was proposed to represent a square grayscale image and resolve the FRQI issue \cite{sang2017novel}. An INEQR (Improved Novel Enhanced Quantum Representation) approach was proposed \cite{su2021improved} that represents tiny rectangular grayscale images rather than a color image. \\
A NASS (normal arbitrary superposition state) was proposed in \cite{li2014multidimensional}. An EFRQI algorithm is proposed in 2021 to minimize the state preparation bits of the NEQR approach\cite{nasr2021efficient}. Grayscale image-based quantum image compression is found in \cite{haque2022advance}. Figure \ref{fig_efrqi} shows an example of EFRQI approach for pixel values of 125(X=0,Y=0), 1(X=1,Y=0), 1(X=4,Y=0), 4(X=0,Y=1), and 16(Y=3,X=0) respectively. For each pixel state preparation, the EFRQI approach uses the same Toffoli gate twice (shown in the red circle) for connecting the state values qubits to the pixel values qubits. To complete the full connection, the Toffoli gate generates a higher amount of bits is the main drawback of the EFRQI approach.  
\section{Proposed methodology}\label{P_M}
\begin{figure}[htbp]
\centerline{\includegraphics[width=\linewidth]{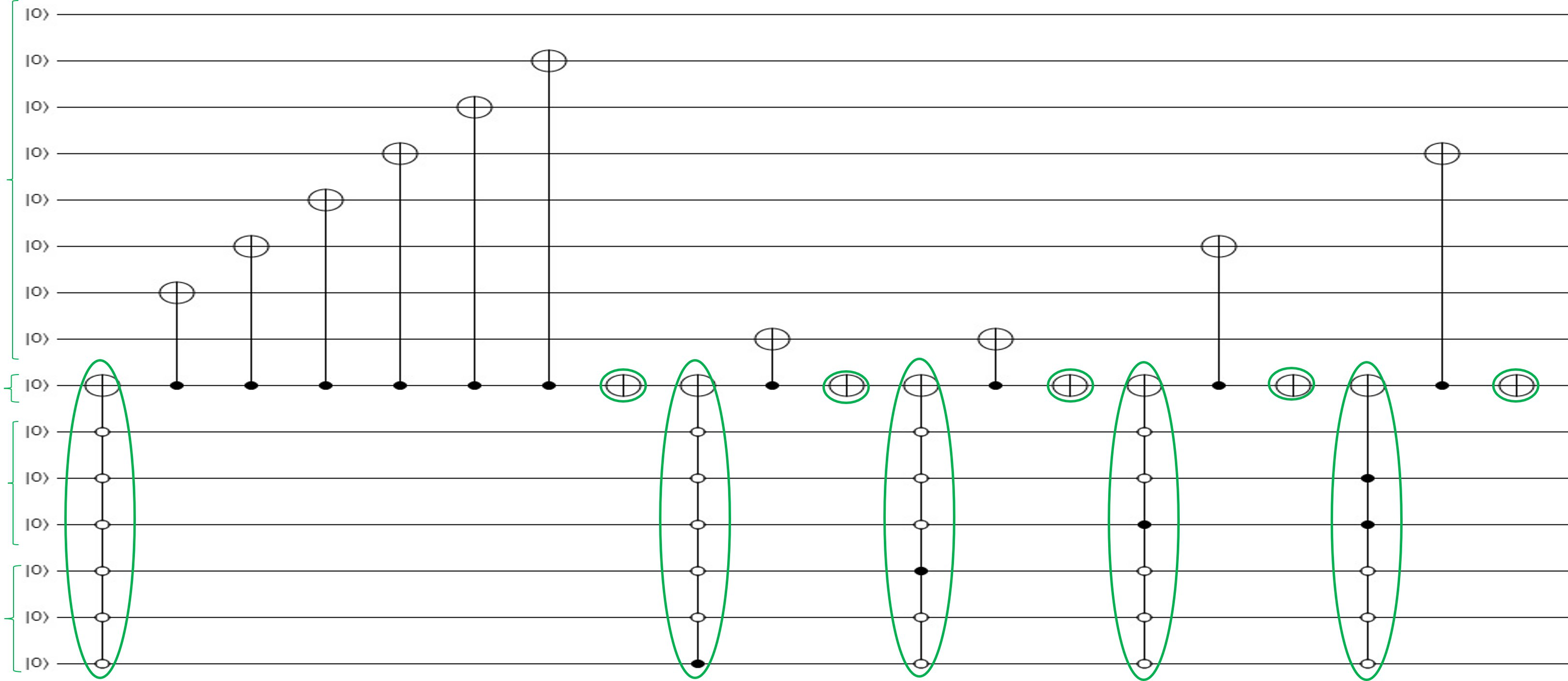}}
\caption{Proposed SCMNEQR circuit diagram.}
\label{fig_proposed}
\end{figure}

Figure \ref{fig_proposed} shows the proposed circuit diagram that represents and compresses the grayscale image. The green circle represents the reset gate used to nullify the previous state’s effect on the next state’s preparation.     
A $2^nX2^m$ image size is considered to represent and compress an image in a block-wise quantum image. Where $m$ and $n$ are the rows and columns of each image.\\ 
The steps involved in the SCMNEQR approach are. \\
Step 1: DCT and quantized. \\
Step 2: Pre-process the pixel or coefficient which requires $q+2n+1$ qubits and initially set all qubit’s values to $\vert0\rangle$. Where $q$ is the number of required qubits to map non-zero pixel or coefficient values. On the other hand, $n_p=log_2(S_x)$  and $m_p=log_2(S_y)$ are the position representing qubits for representing the X and Y position of non-zero pixels or coefficients. Where $S_x$ and $S_y$ are the block size of the X and Y-position images. An auxiliary qubit is used to make the connection between the pixel or coefficient and the position representing qubits. The initial state of qubits can be expressed by the below equation \cite{nasr2021efficient}.
 \begin{equation}
     |\Psi_0\rangle={\vert0\rangle}^{\otimes(q+2n+1)}
 \end{equation}
Then, $(q+1)$ identity gates and $2n$ Hadamard gates are used for pixels or coefficient preparation and its state preparation respectively, and shown below.
\begin{equation}
I=
\begin{bmatrix}
1 & 0 \\
0 & 1 
\end{bmatrix}
\end{equation}
\begin{equation}
H=
\begin{bmatrix}
1/\sqrt2 & 1/\sqrt2 \\
1/\sqrt2 & -1/\sqrt2 
\end{bmatrix}
\end{equation}
In this step, the whole quantum step can be expressed as follows: 
\begin{equation}
U=I^{\otimes{q+1}}\otimes H^{\otimes{2n}}
\end{equation}
The operator $U$ transforms $\Psi_0$ from the initial state to the intermediate state, $\psi_1$.
\begin{equation}
\Psi_1=U(|\Psi_0\rangle)=(I|0\rangle)^{\otimes{q+1}}\otimes (H|0\rangle)^{\otimes{2n}}
\end{equation}
The final preparation step is done using the $U_2$ quantum operator:
\begin{equation}
\Psi_2=U_2(|\Psi_1\rangle)=\frac{1}{2^n} \sum_{i=1}\sum^{j=1}\,|C_{YX}\rangle |YX\rangle
\end{equation}
where $|C_{YX}\rangle$ and $YX$ are pixels or coefficients and the position of the grayscale image. The quantum transform operator is $U_2$ is given below.  
\begin{equation}
U_2=\prod_{X=0,....,2^n-1}\prod_{Y=0,....,2^n-1}\, U_{YX}
\end{equation}
The connection of the Toffoli and reset gate is given below.  
 \begin{equation}
          S_{state}= (log_2(S_X)+log_2(S_Y)+1+1)\otimes{N_{tcn}}
 \end{equation}
 The required bit rate (BR) is calculated using the following equation.
 \begin{equation}
     {BR}=q_{ones}+S_{state}+S_{bit}+A_{bit}+B_e
 \end{equation}
Where $S_{state}$ is the state preparation bit. $q_{ones}$  is the frequent number of ones from pixels or coefficients. $S_{bit}$ is the sign bit that represents the sign of the non-zero pixel or coefficient values. A $N_{tcn}$ is the total number of non-zero coefficient or pixel elements. An $A_{bit}$ is the number of bits that come from the auxiliary qubits. $B_e$ is the bit rate used to locate block position errors.  
Step 3: After applying $8X8$ DCT, extract the non-zero quantized pixel or coefficient values in the $16\times16$ quantum block system. In the meantime, calculate the bit rate (BR) considering with position error of each block. Moreover, the sign bit (SB) is also considered to account for the sign of each non-zero pixel or coefficient value.\\
Step 4: Perform de-quantization. \\
Step 5: Perform inverse DCT. \\
Step 6: Measure the quality of the recovery image.\\
\section{Result and discussion}
\label{R_D}
In this section, the experimental results are analyzed for Deer(1024×1024), Baboons(512×512), Scenery (512×512), and Peppers(512 × 512) images for verification purposes of proposed approach \cite{USC-2022,Washin2022,pan2016new}. Two `experiments have been conducted to demonstrate the proposed method’s performance. Experiments I and II analyze the proposed scheme's computational result directly and indirectly. 
\subsection{Experiment I- result analysis of direct approach}
\begin{figure}[htbp]
\centerline{\includegraphics[width=0.45\linewidth, height=10cm, angle=90]{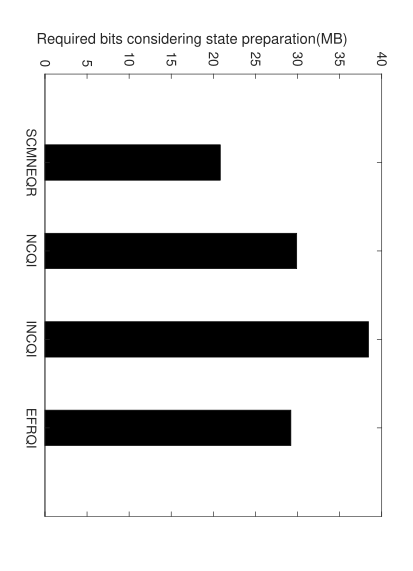}}
\caption{SCMNEQR Bit rate comparison for Deer image}
\label{direct_gray_channel_deer_RDC}
\end{figure}

\begin{figure}[htbp]
\centerline{\includegraphics[width=0.45\linewidth, height=10cm, angle=90]{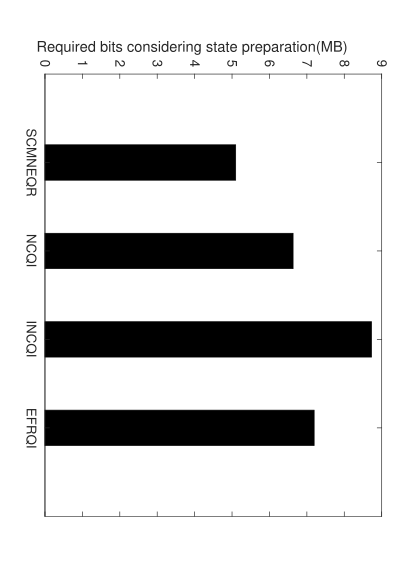}}
\caption{SCMNEQR Bit rate comparison for Baboons image}
\label{direct_gray_channel_baboons_RDC}
\end{figure}

\begin{figure}[htbp]
\centerline{\includegraphics[width=0.45\linewidth, height=10cm, angle=90]{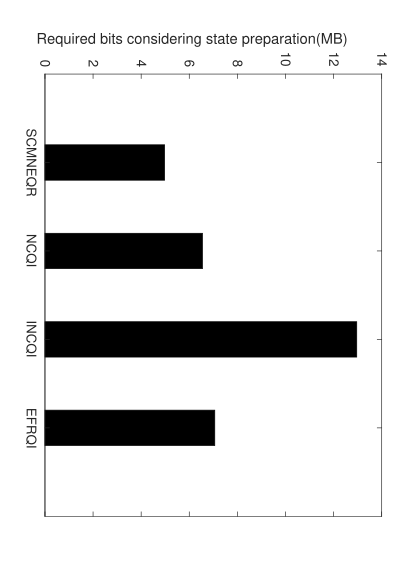}}
\caption{SCMNEQR Bit rate comparison for Scenery image}
\label{direct_gray_channel_scenery_RDC}
\end{figure}

\begin{figure}[htbp]
\centerline{\includegraphics[width=0.45\linewidth, height=10cm, angle=90]{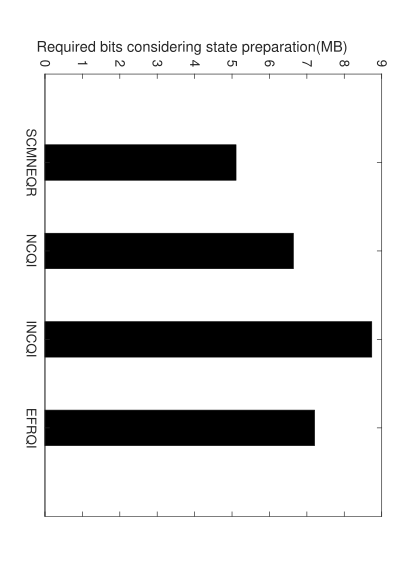}}
\caption{SCMNEQR Bit rate comparison for Pepper image}
\label{direct_gray_channel_peppers_RDC}
\end{figure}

This section analyzes the experimental results related to the SCMNEQR direct approach.  Figure \ref{direct_gray_channel_deer_RDC} exhibits the required bit rate of the proposed SCMNEQR scheme compared with NCQI, INCQI, and EFRQI approaches for deer image. The comparison results show that the proposed SCMNEQR approach requires an approximate (21MB) bit rate which is less compared to NCQI(30MB), INCQI(39MB), and EFRQI(29MB) respectively.\\
On the other hand, Figure \ref{direct_gray_channel_baboons_RDC} shows the required bit rate of the proposed SCMNEQR approach for Baboon's image compared to the other considered approaches. The comparison result reveals that the proposed SCMNEQR approach can represent Baboon's image efficiently because it requires lower bit rates than others. That means, it represents the Baboons image with low complexity circuit. Therefore, the number of required operational gates is lower in the case of the SCMNEQR approach compared to NCQI, INCQI, and EFRQI approaches respectively. 
Figure \ref{direct_gray_channel_scenery_RDC} depicts the required number of bit rates for scenery images and compares the result with NCQI, INCQI, and EFRQI approaches. The comparison result reveals that the SCMNEQR approach requires lower bit rates compared to others. On the contrary, the NCQI and EFRQI require moderate bit rates compared to the SCMNEQR approach.\\
Figure \ref{direct_gray_channel_peppers_RDC} presents the required number of bit rates for the scenery image and compares the result with the existing approaches. Comparison results depict that the SCMNEQR approach represents the image inside the quantum processor more efficiently compared to INCQI, NCQI, and EFRQI approaches respectively. The result is expected due to the efficient architecture of the proposed SCMNEQR approach.
\subsection{Experiment II- result analysis of indirect approach}
\begin{figure}[htbp]
\centerline{\includegraphics[width=\linewidth]{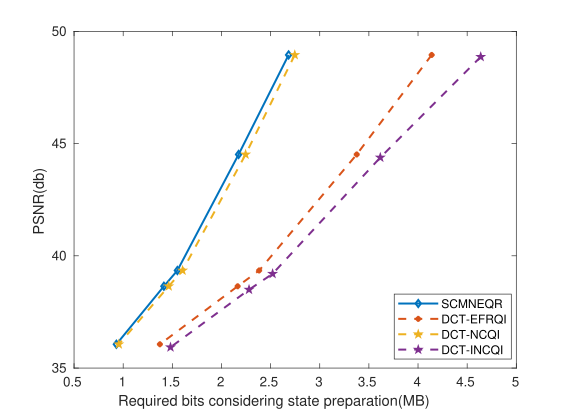}}
\caption{SCMNEQR scheme RDC for Deer image}
\label{gray_channel_deer_RDC}
\end{figure}

\begin{figure}[htbp]
\centerline{\includegraphics[width=\linewidth]{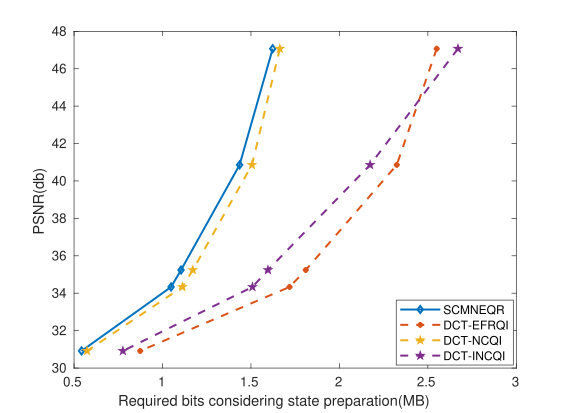}}
\caption{SCMNEQR scheme RDC for Baboons image}
\label{gray_channel_baboons_RDC}
\end{figure}

\begin{figure}[htbp]
\centerline{\includegraphics[width=\linewidth]{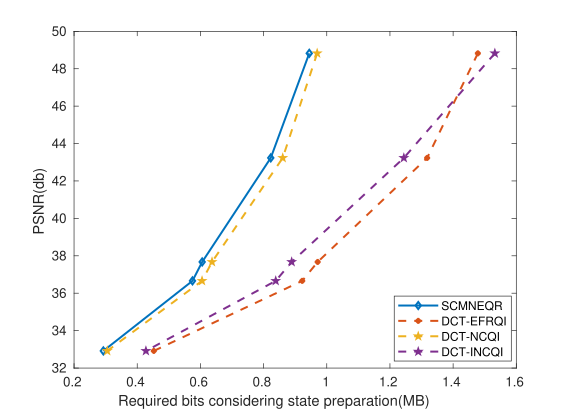}}
\caption{SCMNEQR scheme RDC for Scenery image}
\label{gray_channel_scenery_RDC}
\end{figure}

\begin{figure}[htbp]
\centerline{\includegraphics[width=\linewidth]{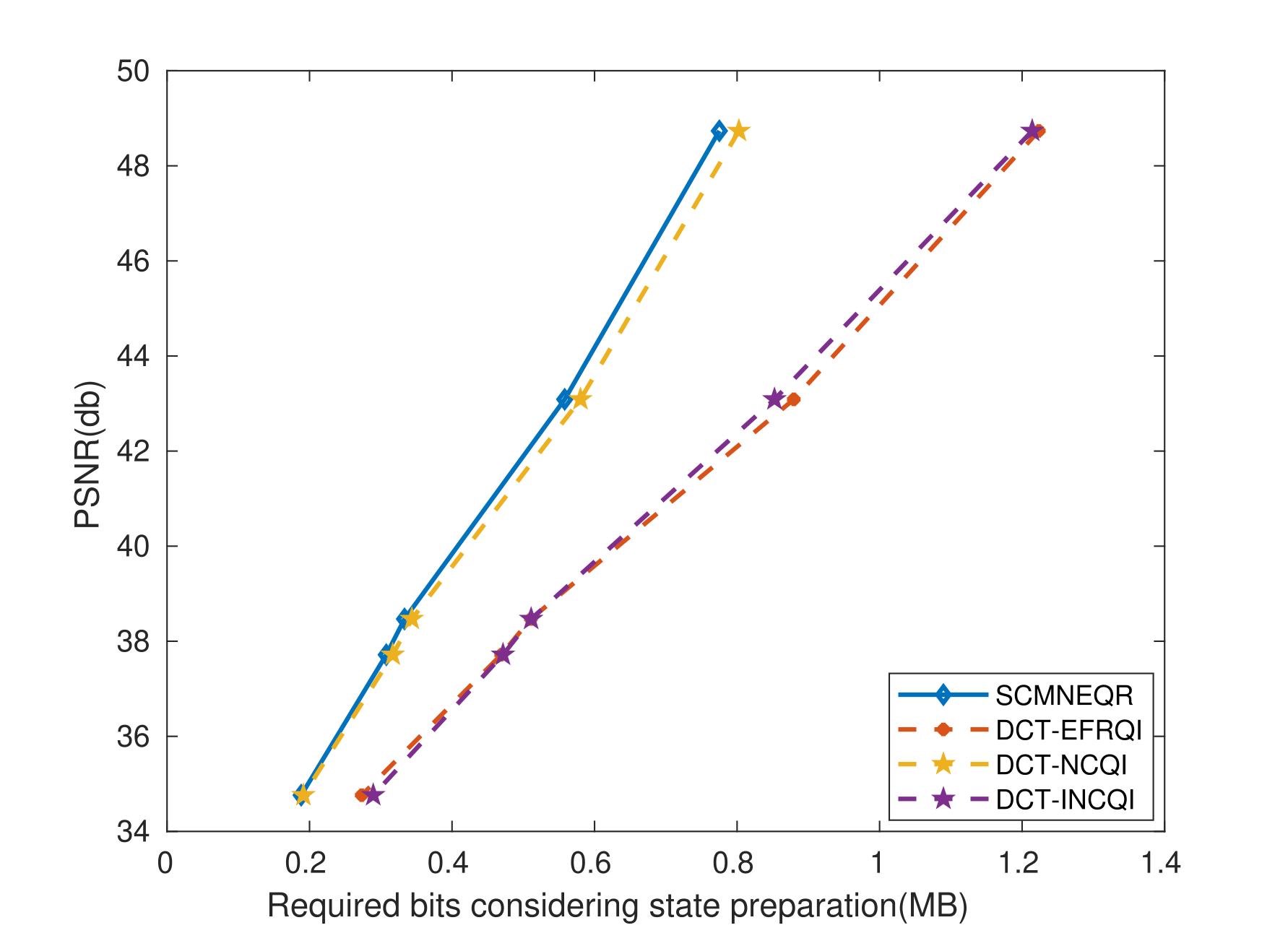}}
\caption{SCMNEQR scheme RDC for Peppers image}
\label{gray_channel_peppers_RDC}
\end{figure}

Figure \ref{gray_channel_deer_RDC} reveals the comparison result of the proposed SCMNEQR approach rate-distortion curve (RDC) compared to DCT-INCQI, DCT-NCQI, and DCT-EFRQI approaches respectively. The comparative result shows that over all the quantization factor, the proposed SCMNEQR approach exhibit a better result compared to others. Meanwhile, only the DCT-NCQI approach draws the near amount of bit rate but the same amount of PSNR values. On the other hand, both DCT-EFRQI and DCT-INCQI require a higher amount of RDC compared to the SCMNEQR approach. From this scenario, It is concluded that the proposed SCMNEQR scheme performs better than represents\ 
Figure \ref{gray_channel_baboons_RDC} presents the comparative results of RDC of the proposed SCMNEQR approach and compares the result with DCT-INCQI, DCT-NCQI, and DCT-EFRQI approaches for Baboon’s image. The comparison result shows that the SCMNEQR approach represents and compresses the gray channel of Baboon’s image efficiently compared to all other considered approaches. The SCMNEQR approach requires lower operational gates than others means that it has more compression ability. Only, DCT-NCQI exhibits an adjacent bit rate whereas DCT-INCQI stays away from the SCMNEQR approach. In between DCT-EFRQI and DCT-INCQI approaches, at the initial quantization factor, QF=8, the DCT-INCQI shows better performance compared to DCT-EFRQI. For QF=16, the DCT-EFRQI requires higher operational gates or bit rates than DCT-INCQI. For this reason, a crossover happened between 8 and 16 quantization factors.  \\
Figure \ref{gray_channel_scenery_RDC} shows the computational results of the SCMNEQR  scheme for scenery images. From a comparative point of view, it has been seen that the SCMNEQR approach has a better ability to compress the scenery image over considered quantization factors compared to all other methods. In between DCT-INCQI and DCT-EFRQI, before QF=16, the DCT-EFRQI approach displays the better result, and thereafter DCT-INCQI performs better over the rest of the other quantization factors. On the contrary, DCT-NCQI requires less bit compared to DCT-INCQI and DCT-EFRQI approaches, but a higher bit rate compared to the SCMNEQR approach.\\ 
Figure \ref{gray_channel_peppers_RDC} shows the computational result of the SCMNEQR
approach alongside DCT-INCQI, DCT-NCQI, and DCT-EFRQI approaches. Results show that the proposed SCMNEQR approach performs better compared to all other considered approaches. The DCT-NCQI approach provides closer RDC compared to the SCMNEQR approach. Conversely, both DCT-INCQI and DCT-EFRQI approaches demonstrate higher RDC values than the SCMNEQR approach. At the quantization factor 8, DCT-INCQI shows better performance compared to the DCT-INCQI approach. After quantization factor 16, the DCT-EFRQI approach requires a higher bit rate than the DCT-INCQI approach. For this reason, a crossover happened between 8 and 16 quantization factors. For the rest of the quantization factor, the DCT-INCQI performs better results compared to the DCT-EFRQI approach.  
\section{conclusion}
\label{CC}
In this article, we have presented a novel quantum circuit for grayscale quantum image representation and compression. The improvement is done in the state connection circuit for efficient representation and compression. One of the major advantages is that it uses a $16\times16$ quantum block. Besides, the required number of qubits for state preparation is 8. Any size of the grayscale image could be presented using the proposed SCMNEQR approach. In addition, it uses the universal quantum Toffoli gate, reset gate, and auxiliary qubit. Another advantage of the proposed scheme is that it does not require a look-up table to perform the operation. It is concluded that the performance of the proposed scheme is much better than the existing approaches.

\section*{Acknowledgment}
The author declare that there is no conflict of interest.

\bibliographystyle{unsrt}  


\end{document}